\documentclass[oneside,11pt,a4paper,english]{article}

\usepackage[utf8]{inputenc}
\usepackage[TS1,T1]{fontenc}
\usepackage[left=2.5cm,right=2.5cm,top=2.5cm,bottom=2.5cm,includefoot]{geometry}
\usepackage[english]{babel}
\usepackage{csquotes}
\usepackage[
    style=science,
    articletitle=true,
    doi=true,
    maxbibnames=99,
]{biblatex}
\usepackage{etoolbox}
\usepackage{mathptmx}
\usepackage{microtype}
\usepackage{siunitx}
\usepackage{physics}
\usepackage{miller}
\usepackage{graphicx}
\usepackage{caption}
\usepackage{amsfonts,amsmath,amssymb}
\usepackage{makecell}
\usepackage{authblk}
\usepackage{hyperxmp}
\usepackage{hyperref}
\hypersetup{
    pdftitle = {Ultrafast nano-imaging of the order parameter in a
    structural phase transition},
    pdfauthor = {Thomas Danz, Till Domröse, Claus Ropers},
}

\newcommand{\tas}{1\textit{T}-TaS$_2$}
\newcommand{\tasbold}{\textbf{1\textit{T}-TaS$_\mathbf2$}}

\title{Ultrafast nano-imaging of the order parameter in a structural phase transition} 
\author[1]{Thomas Danz}
\author[1]{Till Domröse}
\author[1,2,*]{Claus Ropers}

\affil[1]{4th Physical Institute – Solids and Nanostructures, University of Göttingen, 37077 Göttingen, Germany.}
\affil[2]{Max Planck Institute for Biophysical Chemistry, 37077 Göttingen, Germany.}
\affil[*]{Correspondence to: \href{mailto:cropers@gwdg.de}{cropers@gwdg.de}.}
\date{}

\begin{document} 

\maketitle 

\begin{abstract}
\noindent Understanding microscopic processes in materials and devices that can be switched by light requires experimental access to dynamics on nanometer length and femtosecond time scales. Here, we introduce ultrafast dark-field electron microscopy, tailored to map the order parameter across a structural phase transition. We track the evolution of charge-density wave domains in \tas{} after ultrashort laser excitation, elucidating relaxation pathways and domain wall dynamics. The unique benefits of selective contrast enhancement will inspire future beam shaping technology in ultrafast transmission electron microscopy.
\end{abstract}

Optical control over physical and chemical properties of materials is a recurring motif from femtochemistry to ultrafast condensed matter physics. This broad interest is based on the impact of optical control strategies in current and future technology, such as data storage devices \cite{Loke2012,Salinga2013}, neuromorphic computing \cite{Feldmann2019}, photonic circuits \cite{Yang2018}, and energy conversion \cite{Sivula2016}. In parallel, fundamental scientific discoveries reveal unique light-induced properties and correlation effects, involving coupled order parameters \cite{Spaldin2005}, metastable or hidden states \cite{Stojchevska2014,Yoshida2015,Svetin2017,Kogar2020}, superconductivity \cite{Fausti2011}, changes in topology \cite{Sie2019}, and metal-insulator transitions \cite{Morrison2014,Wall2018}.

Typically, microscopic correlations evolve on femto- to picosecond time scales, a regime that is accessible by ultrafast measurement methodology. This approach enables spatially averaged probing of, e.g., electronic gaps \cite{Rohwer2011}, optical conductivity \cite{Falke2014}, magnetization \cite{Stamm2007}, and instabilities against periodic lattice and charge-density modulations \cite{Eichberger2010,Zong2019}. Functionality of devices, however, usually arises from nanoscale structuring or an interplay of different materials. This calls for experimental approaches capturing the dynamics in terms of a spatially dependent order parameter.

Recent experiments accomplished time-resolved mapping of the local free-carrier response in a correlated metal-insulator phase transition using near-field probing \cite{Huber2016,Doenges2016}. Alongside the profound electronic changes observed in these works, direct sensitivity to the spatiotemporal structural modifications governing the transition on the atomic scale remains an open challenge. Ultrafast transmission electron microscopy (UTEM) has proven a valuable tool to study lattice dynamics \cite{Grinolds2006,Baum2007,VanderVeen2013,Piazza2014,Feist2017,Feist2018,Zhang2019a}, but does not provide direct contrast of the structural order parameter.

\begin{figure}[!t]
\centering
\includegraphics[width=\textwidth]{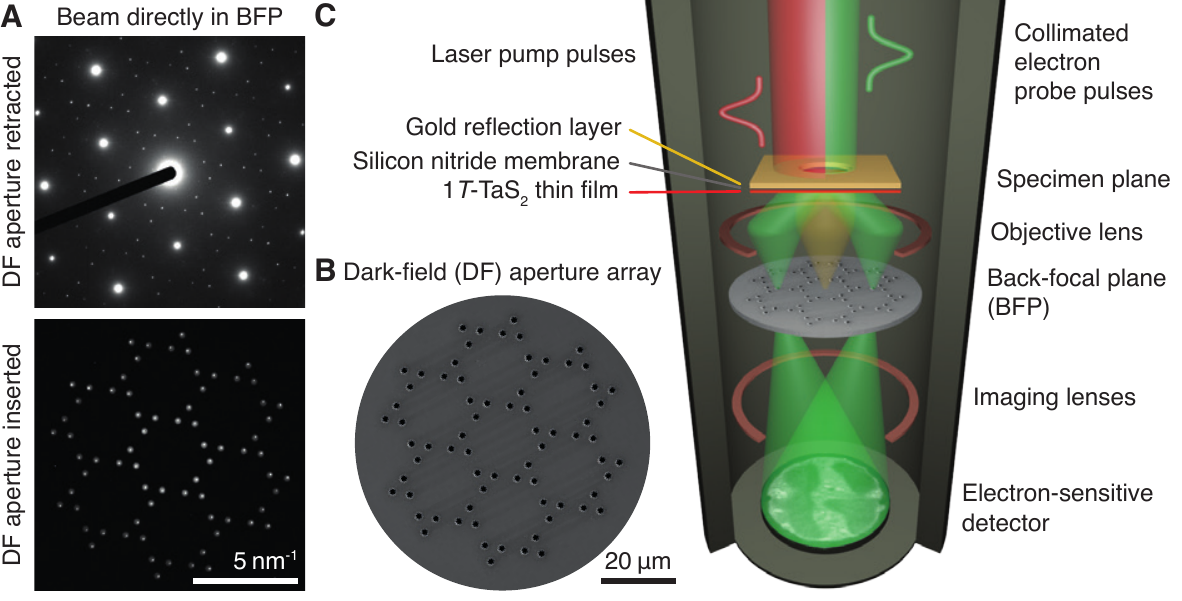}
\caption{
\textbf{Dark-field imaging in the ultrafast transmission electron microscope.}
(\textbf A)~Selection of diffraction spots in the back-focal plane (BFP) using the tailored dark-field (DF) aperture array, as exemplified by electron diffractograms of the \tas{} thin film at room temperature with the DF aperture inserted into (top) and retracted from the electron beam path (bottom). Only superstructure reflections are transmitted through the aperture array.
(\textbf B)~Scanning electron micrograph of the DF aperture array.
(\textbf C)~Sectional drawing of the experimental setup. Electron (green) and optical pulses (red) are incident close to perpendicular on the specimen.
}
\label{fig1}
\end{figure}

In this work, we demonstrate real-space imaging of charge-density wave (CDW) phases with nanometer spatial and femtosecond temporal resolution. Introducing a tailored ultrafast dark-field electron microscopy scheme, we obtain maps of the order parameter in a prototypical CDW system. We observe the formation, stabilization, and relaxation of CDW domains after optical excitation, and identify drastic differences between the emerging domain patterns for continuous-wave and pulsed illumination. Corroborated by time-dependent Ginzburg-Landau simulations, we discuss the non-equilibrium evolution of the order parameter near domain walls.

Specifically, we investigate the transition metal dichalcogenide \tas{}. Owing to its low-dimensional character and strong correlation effects, this system is a promising target for non-trivial order parameter dynamics. It displays several CDW phases with both Mott and Peierls contributions and particularly complex orbital textures \cite{Rossnagel2011,Ritschel2015}. The CDW states and the transitions between them have been the subject of various ultrafast spectroscopy and electron diffraction experiments in the past \cite{Eichberger2010,Sohrt2014,Haupt2016,Laulhe2017,Vogelgesang2018,Zong2018,Storeck2020}.

Our study focuses on the room- and high-temperature phases of the material \cite{Rossnagel2011,Wilson1975,Scruby1975,Spijkerman1997}. Above a temperature of $T^*=\SI{353}{K}$, the material exhibits an incommensurate superstructure aligned with the underlying hexagonal lattice (IC CDW phase; in short IC phase hereafter). Below the phase transition temperature, the CDW transforms into a nearly commensurate superstructure at an angle of \SI{\sim 12}{\degree} between the modulation wave vectors and the lattice directions, effectively reducing the symmetry of the material (NC CDW phase or NC phase). Each of the CDW phases is accompanied by a pronounced periodic lattice distortion (PLD). In electron diffraction images, the PLD of the NC phase is evident from satellite diffraction peaks arranged around the bright structural reflections (see Fig.~\ref{fig1}A, top). Due to a threefold out-of-plane stacking periodicity in the NC phase, mainly second-order satellites are visible \cite{Spijkerman1997}.

In real-space images, NC CDW contrast is obtained by means of dark-field (DF) imaging, i.e., by inserting a DF aperture into the back-focal plane of the microscope’s objective lens. It was recognized early on that DF imaging is in principle suitable to map the spatial distribution of PLDs. However, standard circular DF apertures are too large for this purpose \cite{Wilson1975}, and a single satellite diffraction peak is too weak due to brightness limitations of pulsed electron beams (\SI{<1}{\percent} intensity compared to structural reflections) \cite{Overhauser1971}. Thus, the present work uses focused ion beam etching to manufacture a tailored array of 72 small apertures, selecting the brightest second-order satellite reflections and blocking all other diffraction peaks (see Fig.~\ref{fig1}B) \cite{SM}.

When the DF aperture array is inserted into the column of the electron microscope (see Fig.~\ref{fig1}A, bottom), only electrons scattered into NC satellite diffraction peaks can reach the detector and contribute to the image contrast, while electrons elastically scattered into main lattice peaks and IC diffraction peaks are blocked (see Fig.~\ref{figS1}). Accordingly, the DF images directly reflect the local order parameter $\phi$ of the NC CDW. Kinematical diffraction simulations based on the NC structure \cite{Spijkerman1997} indicate that the dependency between local image intensity and order parameter is close to quadratic with the specific set of reflections considered \cite{SM}. Inelastic scattering contributions account for some background in the image even in the IC phase \cite{Scruby1975}.

\subsection*{Ultrafast dark-field imaging}

\begin{figure}[!p]
\centering
\includegraphics[width=\textwidth]{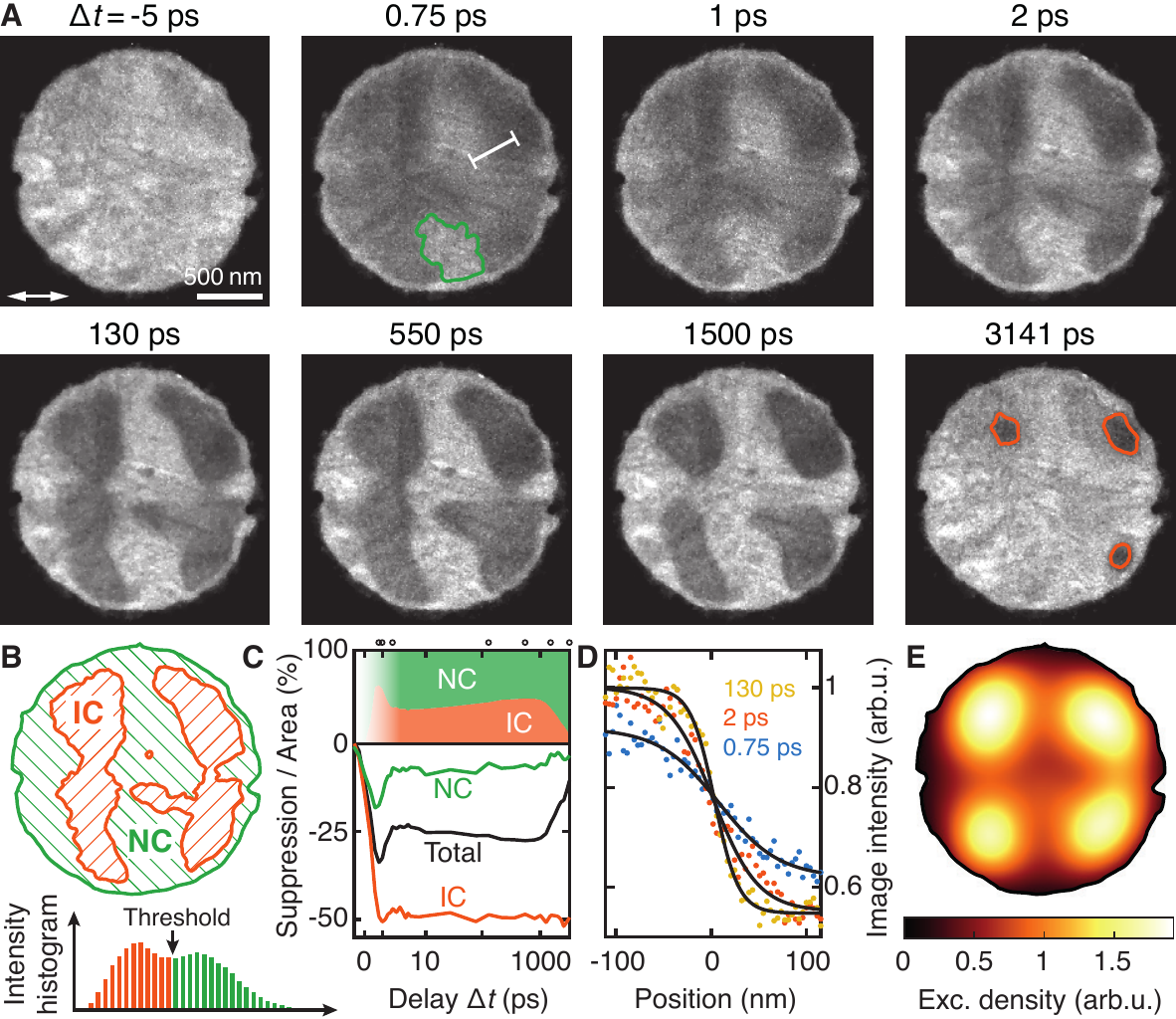}
\caption{
\textbf{Ultrafast dark-field domain imaging of charge-density wave dynamics.}
(\textbf A)~Ultrafast DF micrographs of transient domain configurations in the \tas{} film obtained in the laser pump/electron probe scheme (\SI{2.6}{mJ/cm^2} pump fluence, linear pump polarization indicated by white arrow). Pump/probe delay steps were chosen so as to capture all major stages of the dynamics (see black circles above C).
(\textbf B)~Image segmentation at \SI{130}{ps} delay time. The segmentation threshold is determined from the intensity histogram of the full image series within the circular aperture.
(\textbf C)~Top: Area fractions of NC and IC regions after completed phase separation, as determined from the segmented images. Bottom: Average intensity of the image series within the aperture (black curve), and average intensity in weakly and strongly pumped regions (green/orange curve; evaluated regions are indicated in A).
(\textbf D)~Exemplary profiles of NC/IC phase boundaries taken on the white line indicated in A.
(\textbf E)~Spatial profile of the excitation density giving rise to the initial suppression pattern (see Ref.~\cite{SM} and Fig.~\ref{figS3}).
}
\label{fig2}
\end{figure}

This approach now allows us to directly image CDW dynamics in the material (see Movie~\hyperref[movieS1]{S1}), with a contrast unattainable in conventional bright-field imaging (see Movie~\hyperref[movieS2]{S2}). In our time-resolved experiments, we use femtosecond laser pulses (\SI{800}{nm} center wavelength, \SI{80}{fs} FWHM duration, \SI{420}{kHz} repetition rate) to excite a freestanding \SI{70}{nm} \tas{} film, while probing the transient state of the specimen using ultrashort electron pulses (\SI{120}{keV} electron energy, \SI{530}{fs} FWHM duration). After filtering using the DF aperture array, a spatial image is formed in the detector plane (see Fig.~\ref{fig1}C, and Fig.~\ref{figS2} for more details). Spatially inhomogeneous excitation is required to trigger spatiotemporal dynamics in this experiment. Therefore, we use the circular gold aperture that supports the film as a means to structure the excitation profile by interference between the main beam and edge reflections (see \hyperref[supptext:excitation]{supplementary text} for further information). Additionally, the gold layer acts as a heat bath and prevents thermal load on the specimen outside the region of interest \cite{SM}. This very local excitation and effective thermal coupling ensures reversibility of the observed dynamics at an unprecedented repetition rate for a structural phase transition.

Ultrafast DF images are shown in Fig.~\ref{fig2}A as a function of optical pump/electron probe delay $\Delta t$ \cite{SM}. Initially, the image within the circular aperture is flat with minor diffraction features and spatial variations of the NC CDW/PLD amplitude at room temperature. Directly after the temporal overlap of electron and laser pulses (‘time-zero’), an inhomogeneous suppression of image intensity is evident with its shape governed by the excitation profile (see Fig.~\ref{fig2}E). Within two picoseconds, well-defined domains have emerged, and the boundaries separating bright from dark regions have become visibly sharper (see Fig.~\ref{fig2}D). On a \SI{100}{ps} time scale, a slight growth of the darkened regions and a further contrast sharpening is observed. Finally, these domains diminish in area, and a homogeneous NC contrast is re-established after a few nanoseconds.

The image series contains considerably more information than spatially averaged diffraction data alone (cf. Fig.~\ref{figS4}). Specifically, we can individually analyze the PLD dynamics in weakly and strongly pumped regions, which either show a transient quench and recovery, or a persistent phase transformation with a full suppression within the temporal resolution (green and orange curves in Fig.~\ref{fig2}C, bottom) \cite{Eichberger2010}. The practically binary contrast is a clear sign of a completed phase separation into distinct NC and IC regions no later than few picoseconds after the pump. This allows us to carry out image segmentation in order to identify the temporal evolution of the respective area fractions (see Figs.~\ref{fig2}B and \ref{figS5}). In particular, this analysis shows that an ongoing suppression of the spatially averaged contrast (black curve in Fig.~\ref{fig2}C), starting after a few picoseconds and lasting up to one nanosecond, is caused by a growth of the IC domains (see moderate increase of IC fraction in Fig.~\ref{fig2}C, top).

\clearpage
\subsection*{Steady-state domain imaging}

\begin{figure}[!b]
\centering
\includegraphics[width=\textwidth]{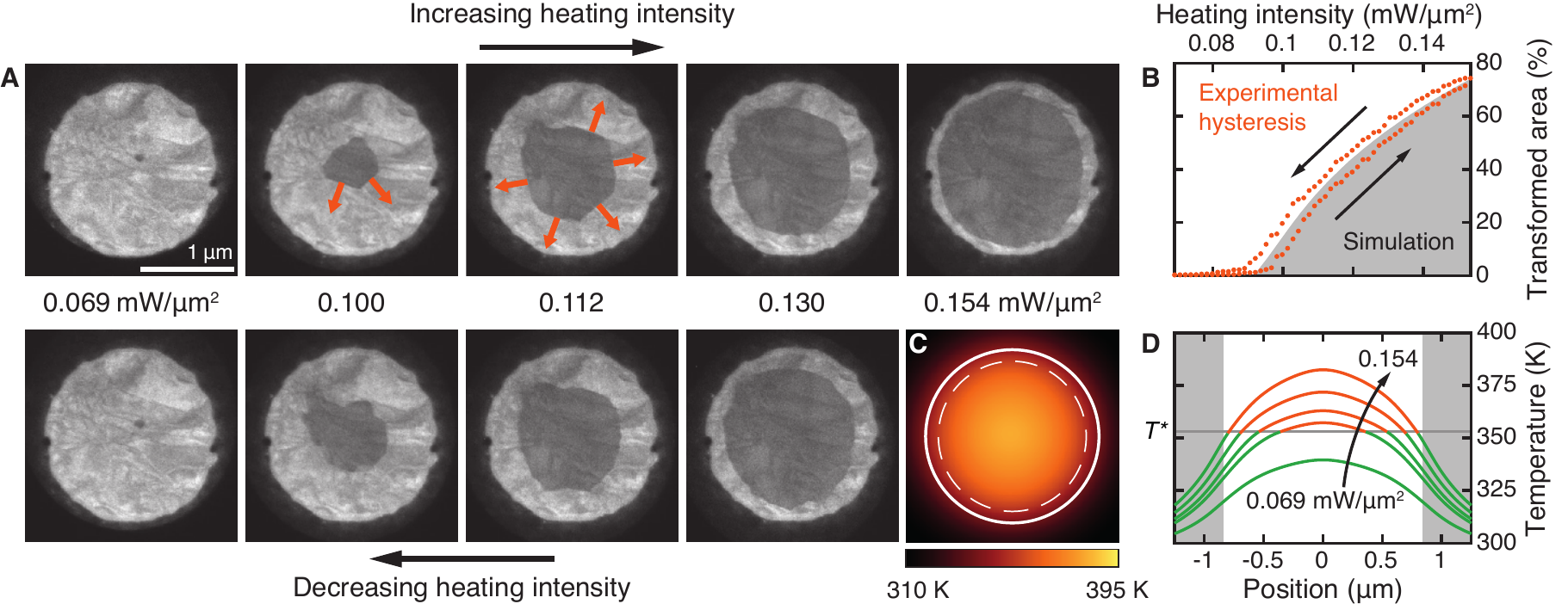}
\caption{
\textbf{Dark-field domain imaging with continuous-wave laser excitation.}
(\textbf A)~DF micrographs of laser-intensity up-sweep and down-sweep. Orange arrows indicate phase boundaries oriented along preferential lattice directions.
(\textbf B)~Hysteretic behavior as extracted from the micrographs (orange dots). The characteristic intensity dependency is reproduced by heat transfer simulations (gray).
(\textbf C)~Temperature distribution simulated for a laser heating intensity of \SI{0.154(9)}{mW/\micro m^2} for comparison with the experimental result (solid line: edge of circular aperture; dashed line: edge of transformed area).
(\textbf D)~Radial profiles of the simulated temperature distribution in the film for the experimental laser intensities presented in A. Green (orange) segment of the curves: temperature below (above) the phase transition temperature $T^*$. Shaded regions are outside the aperture.
}
\label{fig3}
\end{figure}

The CDW phase pattern in the time-resolved experiment is a direct consequence of the spatiotemporal non-equilibrium, and strongly deviates from the situation observed under constant heating of the structure using a \SI{532}{nm} continuous-wave laser (see Fig.~\ref{fig3}A and Movie~\hyperref[movieS3]{S3}): Upon increasing the laser heating intensity, an IC domain nucleates close to the center of the thin film, and grows until it almost fills the circular aperture. Decreasing the laser intensity results in a reduction of the switched area.

The steady-state phase pattern arises from the dynamical equilibrium between deposited and dissipated energy. Due to efficient equilibration within the film compared to the thermal coupling to the support and heat bath, in this experiment, the optical excitation profile is not evident. This provides us with a detailed picture of the sample’s thermal state. We use a heat transfer simulation to predict the temperature distribution for each laser intensity (see Fig.~\ref{fig3}, B-D), and quantify the thermal boundary resistance between the film and the substrate. The value obtained falls well in line with those of related interfaces \cite{SM}.

A slight hysteresis of \SI{<4}{K} is observed in the heating cycle (see orange dots in Fig.~\ref{fig3}B), a signature of the first-order character of the phase transition \cite{Wilson1975}. On microscopic length scales, we identify two different mechanisms affecting the hysteretic behavior: The up-sweep reveals a preferential formation of NC/IC phase boundaries along specific crystallographic directions (indicated by orange arrows in Fig.~\ref{fig3}A; see \hyperref[supptext:orientation]{supplementary text} for details). On the other hand, local pinning sites visibly affect the down-sweep due to some degree of structural inhomogeneity (cf. Fig.~\ref{figS6}, A-C).

The appearance of these effects only under continuous excitation is a result of very different energy and time scales governing the dynamics. The steady-state experiment is rather sensitive to local defects and structural anisotropy, as there is sufficient time for small spatial variations in the CDW free-energy landscape to determine the phase pattern. On the other hand, the time-resolved experiment is characterized by electron-lattice non-equilibrium involving strong thermal gradients.

\subsection*{Simulation of CDW dynamics}

\begin{figure}[!b]
\centering
\includegraphics[width=\textwidth]{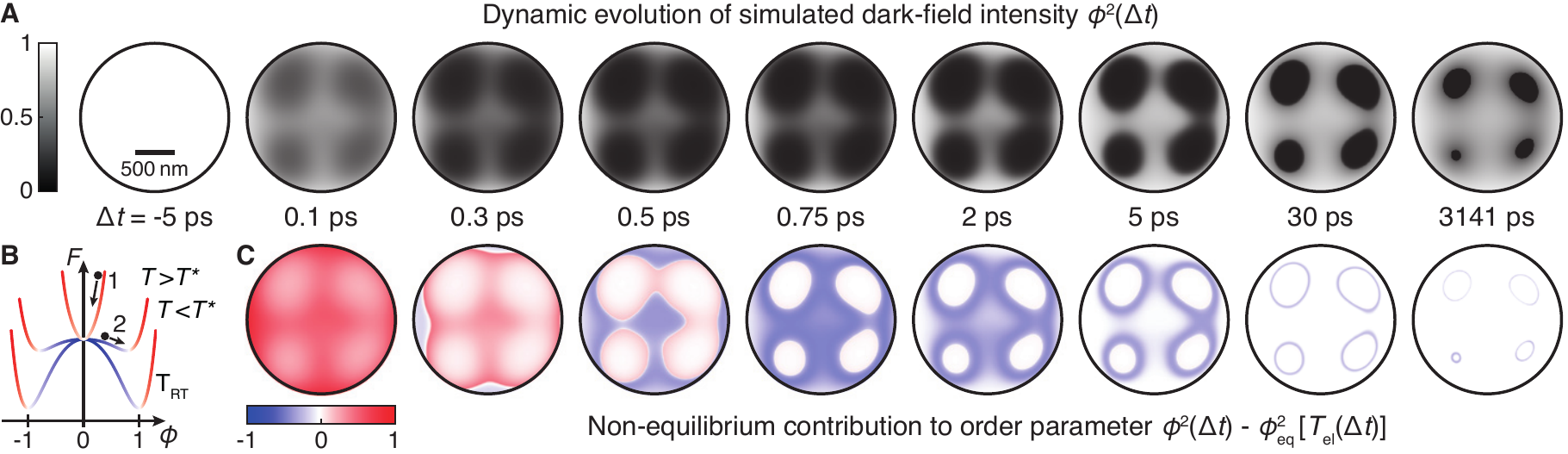}
\caption{
\textbf{Time-resolved Ginzburg-Landau simulations of optically induced order parameter dynamics.}
(\textbf A)~Series of simulated dark-field micrographs (squared order parameter $\phi^2$ projected along the specimen thickness in order to mimic DF image contrast).
(\textbf B)~Sketch of free-energy landscapes for different electron temperatures (color scale as described in C).
(\textbf C)~Deviation of the squared order parameter from its local equilibrium value $\phi_\text{eq}^2$ (red/blue color: order parameter is larger/smaller than at the free-energy minimum).
}
\label{fig4}
\end{figure}

In order to obtain a better understanding of the spatiotemporal CDW evolution and energy redistribution, we simulate the three-dimensional dynamics of the order parameter as well as the electron and lattice temperatures. Employing a Ginzburg-Landau model for a non-conserved order parameter, we capture the symmetry-breaking nature of the first-order NC/IC phase transition (see Ref.~\cite{Hohenberg1977}, model ‘A’). As in the experiments, we do not separately analyze the individual wave vector components of the CDW. Accordingly, we consider the spatially varying PLD amplitude as a single, real-valued order parameter $\phi$ with a generic free-energy functional for a first-order phase transition, parametrized by the local electron temperature $T_\text{el}$. The evolution of the electron and lattice temperatures is treated with coupled heat diffusion equations using the thermal parameters extracted from the steady-state experiment \cite{SM}.

For values of $T_\text{el}$ below the phase transition temperature $T^*$, the free-energy landscape exhibits a double-well shape (see Fig.~\ref{fig4}B). Initially, the simulation volume is in thermodynamical equilibrium with $\phi=1$, corresponding to one of the potential minima at room temperature. The optical excitation at time-zero is modeled as a jump in electronic temperature with the spatial distribution given by the excitation profile (see Fig.~\ref{fig2}E). In this highly non-equilibrium state with $T_\text{el} > T^*$ in large areas of the sample, the reshaped potential is close to parabolic with a single IC CDW minimum at $\phi=0$. The subsequent relaxation of the order parameter towards the high-symmetry state causes a global order parameter suppression at early times. Following equilibration of electron and lattice temperatures on a time scale of \SI{350}{fs} \cite{Eichberger2010}, the simulation in Fig.~\ref{fig4}A and Movie~\hyperref[movieS4]{S4} closely reproduces the phase separation and cooling observed in the experimental dataset (cf. Fig.~\ref{fig2}A).

Local trajectories of the order parameter can be classified according to the magnitude of its non-equi\-lib\-ri\-um contribution (see Fig.~\ref{fig4}C and Movie~\hyperref[movieS5]{S5}). In regions with sufficient excitation density, the order parameter relaxes into the IC potential minimum within the electron-lattice coupling time (see point ‘1’ in Fig.~\ref{fig4}B). The spatiotemporal evolution of the corresponding four white patches adiabatically follows the reshaping of the potential on heat diffusion time scales. On the other hand, and as a consequence of the nearly flat potential just below the phase transition temperature, more weakly excited regions undergo a transient quench of the NC phase, relaxing on a few-picosecond time scale (point ‘2’) (see also Fig.~\ref{figS7}).

Close to the phase boundaries, the suppression of NC charge-ordering lasts until \SI{30}{ps} after excitation (see blue contours forming around the white patches after \SI{2}{ps}), and a minor slowdown of the IC formation is observed as well (light red regions at \SI{0.5}{ps}). These processes are a direct consequence of the delicate interplay between fast electron-lattice coupling and order parameter dynamics in a reshaped free-energy landscape, and ultimately result in the sharpening of domain walls observed in simulation and experiment (see Fig.~\ref{figS8}).

\subsection*{Conclusions}

In conclusion, this study reveals insights into the non-equilibrium dynamics of a nanoscale phase transition on femtosecond timescales and with \SI{5}{nm} spatial resolution \cite{SM}, harnessing previously unattainable experimental contrast. Aided by a precise steady-state characterization of the sample, we demonstrate that the phase pattern and, specifically, the dynamics at phase boundaries can be described by a theoretical model with a minimal set of macroscopic assumptions on the symmetry and thermodynamics of the phase transition.

Our study illustrates how contrast enhancements by post-specimen \cite{Danev2010,Verbeeck2010,Grillo2017,Schwartz2019} or even pre-specimen beam shaping \cite{Ophus2016} allow for sensitivity to further degrees of freedom in complex materials, paving the way for new types of investigation. These possibilities include imaging of transient phonon populations \cite{Stern2018}, and particularly visualizing amplitude and phase modes in charge-density wave materials \cite{Overhauser1971}. Furthermore, spatially resolved investigations of topological defects in the charge-ordering of these materials may lead to a deeper understanding of phase formation kinetics \cite{Zong2019,Haupt2016,Laulhe2017,Vogelgesang2018}. Lastly, careful design of the beam shaping parameters may facilitate multi-phase imaging, or the simultaneous mapping of chiral mirror domains \cite{Zong2018}. Building upon these opportunities, ultrafast transmission electron microscopy will offer extraordinary perspectives for observing nanoscale dynamics in correlated materials.

\apptocmd{\sloppy}{\hbadness 10000\relax}{}{}
\sloppy{\printbibliography[title={\large References}]}

\subsection*{Acknowledgments}
The authors thank Murat Sivis for technical support in focused ion beam milling of specimen and DF aperture array, as well as Kai Rossnagel (University of Kiel) for supplying high quality \tas{} crystals. We acknowledge assistance from the Göttingen UTEM team and especially Tyler Harvey who designed the excitation beam path. Furthermore, we are grateful to Carolin Wichmann for providing the in-plane rotation specimen holder.

\subsection*{Funding}
This work was funded by the Deutsche Forschungsgemeinschaft (DFG) in the Collaborative Research Center “Atomic scale control of energy conversion” (DFG-SFB 1073, project A05) and via resources from the Gottfried Wilhelm Leibniz prize. Th.D. gratefully acknowledges a scholarship by the German Academic Scholarship Foundation.

\subsection*{Author contributions}
Th.D. and Ti.D. conducted the experiments and analyzed the data. Th.D. conceived and manufactured the DF aperture array. Th.D. and Ti.D. prepared the specimen and implemented the time-resolved simulations. Th.D. carried out the steady-state simulations and wrote the manuscript with contributions from all authors. C.R. conceived and directed the study. All authors discussed the results and their interpretation.

\subsection*{Competing interests}
The authors declare no competing interests.

\subsection*{Data and materials availability}
The data that support the findings of this study are available from the corresponding author upon request.

\clearpage
\section*{Supplementary materials}

\renewcommand{\thefigure}{S\arabic{figure}}
\setcounter{figure}{0}
\renewcommand{\thetable}{S\arabic{table}}
\setcounter{table}{0}

\subsection*{Materials and methods}

\paragraph*{Ultrafast transmission electron microscopy in DF mode}

The Göttingen Ultrafast Transmission Electron Microscope (UTEM) is based on a commercial JEOL JEM-2100F Schottky field emission microscope, modified to allow for ultrashort electron pulse generation from a ZrO/W Schottky field emitter tip using \SI{400}{nm} ultrashort laser pulses. This high-coherence ultrafast electron source enables the investigation of spatiotemporal dynamics in a laser pump/electron probe scheme with sub-nanometer spatial resolution, \SI{0.6}{eV} energy resolution, and down to \SI{200}{fs} temporal resolution. Technical details on the instrument are given in Ref.~\cite{Feist2017}.

In this study, the \SI{800}{nm} pump pulses arrive at the specimen near normal incidence (\SI{\sim 6}{\degree}) and are focused down to \SI{15}{\micro m} FWHM using an aspherical lens. A collimated electron beam illuminates the specimen with a \SI{2.4}{\micro m} spot diameter (slightly larger than the \SI{1.8}{\micro m} circular gold aperture supporting the membrane; see subsection “\nameref{mat:specimen}”). The specimen is mounted in an in-plane rotation specimen holder (Fischione Model 2040).

In DF mode, the DF aperture array filters NC CDW wave vector components in the back-focal plane of the objective lens where a first electron diffraction pattern is formed (see extended schematic in Fig.~\ref{figS2}A). Finally, the real-space image is captured either on a direct detection camera (Direct Electron DE-16 in “counting mode”), or a conventional scintillator-coupled CCD (Gatan UltraScan 4000). The CCD detector has only been used in the steady-state experiment (see Fig.~\ref{fig3}A and Movie~\hyperref[movieS3]{S3}).

\paragraph*{Manufacturing of DF aperture array and theoretical resolution limit}

In order to introduce the DF aperture array into the microscope, we constructed an objective lens aperture holder that accommodates two standard-sized \SI{3}{mm} silicon nitride grids for transmission electron microscopy (TEM) (see Fig.~\ref{figS2}, B and C). Each TEM grid (Silson; \SI{50}{nm} Si$_3$N$_4$ film thickness, \SI{500}{\micro m} window size) is covered with a polycrystalline, sufficiently electron-opaque gold film of \SI{600}{nm} nominal thickness by magnetron sputtering. Subsequently, we prepared the aperture array (see Fig.~\ref{figS2}D) using focused ion beam milling (FEI Nova NanoLab 600 DualBeam).

Based on Fig.~\ref{figS2}E, we determine an actual gold film thickness of \SI{780}{nm} and a hole diameter of \SI{1.8}{\micro m}. Given the corresponding passband in reciprocal space, we estimate an achievable spatial image resolution in DF mode better than \SI{5}{nm}.

\paragraph*{Specimen preparation and characterization}
\label{mat:specimen}

We used magnetron sputtering to deposit a \SI{5}{nm} titanium adhesion layer and a \SI{200}{nm} gold film on the backside of standard-sized \SI{3}{mm} silicon nitride TEM grids (Norcada; \SI{30}{nm} Si$_3$N$_4$ film thickness, \SI{10}{\micro m} window size). Afterwards, we created a \SI{1.85}{\micro m} circular through-hole in the center of the \SI{10}{\micro m} window by focused ion beam milling, and deposited a \tas{} flake obtained by ultramicrotomy on top (Leica Ultracut UCT with DiATOME Ultra \SI{45}{\degree} diamond knife; \SI{50}{nm} nominal thickness) \cite{Danz2016}.

Figure~\ref{figS6}, A-C shows electron micrographs of the specimen structure obtained using different imaging modes with a continuous electron beam. In order to characterize the actual specimen thickness, we use STEM-EELS (scanning transmission electron microscopy with an electron energy loss spectrum recorded per image pixel) at \SI{200}{kV} acceleration voltage (\SI{\sim 13.0}{mrad} convergence angle, \SI{\sim 14.3}{mrad} collection angle), and extract the local specimen thickness per pixel in units of the electron mean free path (MFP) \cite{Egerton2011}. Using the formula derived by Iakoubovskii \textit{et al.} \cite{Iakoubovskii2008}, we calculate the MFP of \tas{}, gold, and the silicon nitride membrane based on the convergence angle, the collection angle, and the respective material densities (neglecting the thin titanium adhesion layer) \cite{Haynes2017,Norcada2017}. This results in an MFP of \SI{126}{nm} for \tas{}, \SI{153}{nm} for the silicon nitride membrane, and \SI{99}{nm} for gold. Treating the \SI{30}{nm} silicon nitride thickness as fixed, we obtain actual layer thicknesses of \SI{70}{nm} for the free-standing \tas{} flake, and \SI{145}{nm} for the gold film (averaged along the image edges) (see Fig.~\ref{figS6}, D and E).

\paragraph*{Image post-processing, image segmentation, and delay curves}

Due to limited maximum exposure times of the cameras, each individual image taken consists of a certain number of frames (between one and 60, depending on the specific experiment) that are added up to obtain the final image. For data taken with the direct detection camera under low-dose conditions and presented in the main text, effective integration times per image range between \SI{11}{min} (see Fig.~\ref{fig2}C) and \SI{65}{min} (see Fig.~\ref{fig2}A). Fluctuations in beam current over time are compensated by image normalization based on the intensity of reference images taken at regular intervals and under identical experimental conditions (before time-zero).

After removal of fixed-pattern noise, we use total variation denoising to reduce the presence of shot noise in the experimental image series while retaining edges \cite{Biguri2016}, especially those between regions of different CDW phases. Additionally, we align the individual images in each image series using the edge of the circular gold aperture as a reference in order to compensate for potential specimen drift during the experiment.

An image segmentation approach is applied to convert the NC/IC CDW image contrast to binary masks of the two phases, obtaining their respective occupied areas. In order to calculate the image intensity threshold of the full image series, we use Otsu’s method which minimizes the intra-class variance and maximizes the inter-class variance of bright and dark pixels \cite{Otsu1979}. After applying the threshold to the image series, we remove NC and IC regions from the binary masks whose area is smaller than a certain threshold in order to reduce fragmentation of the masks.

Finally, delay curves for weakly and strongly pumped regions (see green/orange curve in Fig.~\ref{fig2}C) are derived by spatially averaging the image intensity over the regions indicated in Fig.~\ref{fig2}A at 0.75 and \SI{3141}{ps}. The average signal (black curve) is obtained from the total counts inside the circular aperture. All logarithmic delay axes in this work are generated using a symmetric logarithm transformation in order to be able to include data before and close to time-zero as well \cite{Webber2013}.

\paragraph*{Extraction of excitation profile from experimental data}
\label{mat:excitation}

The spatial profile of the excitation density shown in Fig.~\ref{fig2}E is extracted from an ultrafast DF image series recorded as a function of pump fluence at a fixed pump/probe delay of \SI{0.75}{ps} (see Fig.~\ref{figS3}). After image segmentation, we assign the peak excitation density $I_1$ to the region that is being transformed into the IC phase already at the lowest pump fluence $F_1$. For all areas of the specimen transformed into the IC phase at a fluence $F_i$, we extract an excitation density $I_i=I_1\cdot F_1/F_i$. The final excitation profile is obtained after Gaussian smoothing and normalizing to the average within the gold aperture.

\paragraph*{Kinematical diffraction simulation of the NC phase}

The presence of the nearly commensurate CDW in the NC phase of \tas{} leads to superlattice reflections that appear in the diffraction pattern in addition to the structural Bragg reflections. Overhauser described the structure factor for a sinusoidal charge-density modulation in a linear chain of atoms in Ref.~\cite{Overhauser1971}. However, for a three-dimensional, nearly commensurate CDW as in \tas{}, and in the presence of possible higher-harmonic contributions to the charge-density modulation, an analytic calculation of structure factors becomes much more difficult.

In order to compare experimental data with simulated order parameter dynamics, we need access to the dependency between the order parameter of the NC phase (the CDW/PLD amplitude) and the total intensity of the 72 superlattice reflections that contribute to the DF image contrast. We use a dataset of atom positions by Spijkerman \textit{et al.} which includes a number of different harmonics to describe the modulation of both tantalum and sulfur positions in the material \cite{Spijkerman1997}. The atomic scattering factors of tantalum and sulfur are included in the calculations \cite{Colliex2006}. Additionally, we approximate the NC CDW structure by a commensurate one with a size of $147\times 147\times 3$ undistorted unit cells.

In this setting, we calculate the structure factors of the 72 reflections as a function of CDW amplitude by linearly scaling the harmonic coefficients between zero and their full value. A fit of CDW amplitude vs. simulated DF image intensity reveals a power law scaling with an exponent of approximately 1.92. For the sake of simplicity, we use a value of 2 in the main text and Fig.~\ref{fig4}.

\paragraph*{Finite element simulations: General setup}

The simulation results outlined in the main text are obtained using finite element simulations in COMSOL Multiphysics 5.4. All simulations are conducted using the same model of the specimen but varying meshes and sizes of the simulation volume according to the requirements of each simulation step.

The lateral dimensions of the full simulation geometry cover the area of the \SI{10}{\micro m} silicon nitride window (see Fig.~\ref{figS6}F). Outside of the window, sufficient thermal coupling to the heat bath/\SI{200}{\micro m} thick silicon frame of the TEM grid is assumed. \tas{} and gold layers of equal size are positioned on the bottom and top of the silicon nitride membrane, respectively. Gold and silicon nitride material is removed within the circular aperture. We neglect the presence of the thin titanium adhesion layer in the model. All layer thicknesses and geometry dimensions are defined as determined from the actual specimen (see subsection “\nameref{mat:specimen}”). Throughout the different simulations, we use material properties taken from the references listed in Table \ref{tabS1}. For the thermal properties, we use temperature-dependent values where possible, and constant extrapolation outside the available data range. The heat bath is at room temperature at all times ($T_\text{RT} = \SI{293.15}{K}$).

\paragraph*{Finite element simulations: Field calculations}
\label{mat:fields}

Initially, we calculate the absorption behavior for both optical pump wavelengths in this study (\SI{532}{nm} and \SI{800}{nm}) using the “Wave Optics” module of COMSOL in order to derive the absorbed laser power per unit volume at any point of the specimen. In order to reduce computational complexity, we only calculate the optical fields inside and close to the circular aperture (two wavelengths from aperture edge to simulation boundary). Additionally, we harness the symmetry of the model by reducing the simulation volume to one quadrant of the geometry and setting appropriate boundary conditions for the field (“Perfect Magnetic Conductor” at stitch boundaries where the electric field is tangential to the boundary, and “Perfect Electric Conductor” where the electric field is normal to the boundary).

The simulation itself is carried out using a scattered field formulation for the electric field. In a first step, we calculate the fields resulting from a plane wave impinging on the specimen stack in perpendicular incidence with the circular aperture removed (“background field”). In the second step, we calculate the “scattered field” of the full structure including the circular aperture, taking the background field into account. The sum of background field and scattered field then gives a precise solution for the field distribution inside and around the structure.

\paragraph*{Finite element simulations: Steady-state experiment}

We simulate the domain pattern of the steady-state experiment using the “Heat Transfer in Solids” module of COMSOL. We now take the full size of the simulation geometry into account, because we can only assume sufficient thermal coupling to the heat bath at the edges of the silicon nitride window. However, we are still able to reduce the model to a single quadrant due to symmetry (using “Thermal insulation” boundary conditions at the stitch boundaries). The absorbed laser power per unit volume at a pump wavelength of \SI{532}{nm} (see \hyperref[mat:fields]{previous subsection}) is entered as a three-dimensional heat source. Due to the layered nature of \tas{}, we assume a reduction of thermal conductivity by a factor of $\num{\sim 3}$ perpendicular to the layers \cite{Hellmann2012}. Then we use both the incident laser intensity and the thermal conductivity of the silicon nitride membrane to fit the simulation results for the transformed area vs. incident laser intensity curve to the experimental data (see Fig.~\ref{fig3}B).

The obtained value for the heat conductivity of the silicon nitride layer amounts to \SI{0.9(2)}{\percent} of the silicon nitride bulk value; thus, it is dominated by the quality of the \tas{}/silicon nitride and silicon nitride/gold interfaces. From this, we deduce a total interfacial thermal resistance of \SI{\sim 0.7e-6}{m^2 K/W}, which falls well in line with results for the thermal contact between a MoS$_2$ flake and platinum electrodes \cite{Aiyiti2018}. From the fit, we obtain a reasonable value of \SI{20}{\micro m} FWHM for the focused laser spot size at \SI{532}{nm}.

\paragraph*{Finite element simulations: Heat transfer in ultrafast experiment}

Based on the specimen model as refined in the steady-state experiment, we simulate the time-dependent heat transfer underlying the specimen response in the ultrafast experiment. We describe the non-equilibrium state of the \tas{} layer in the first few picoseconds using a two-temperature approach for both electron and lattice systems, while assuming full electron-lattice equilibration at all times in gold and silicon nitride.

We consider a linear electronic heat capacity $C_\text{el} =\gamma\cdot T_\text{el}$ for the \tas{} layer with a linear coefficient of $\gamma=\SI{8.5}{mJ/mol/K^2}$ (cf. Ref.~\cite{Meyer1975}), and we choose the lattice heat capacity $C_\text{ph}$ such that we retain $C_\text{total} = C_\text{el} + C_\text{ph}$ as given in Ref.~\cite{Suzuki1985}. Electron and lattice thermal conductivities are distributed between both subsystems as given in Ref.~\cite{Nunez-Regueiro1985}. The electron-lattice coupling constant is selected such that subsystems equilibrate on a timescale of $\tau_\text{el-ph} = \SI{350}{fs}$ \cite{Eichberger2010}.

The COMSOL field calculations do not closely reproduce the observed excitation pattern due to the precise shape of the actual gold aperture in the experiments. Instead, we combine the depth dependency from the field calculations with the spatial profile of the excitation density as extracted from experimental data (shown in Fig.~\ref{fig2}E; see subsection “\nameref{mat:excitation}”). This excitation pattern breaks the four-fold symmetry of the model; thus, we simulate the full $\SI{10}{\micro m}\cdot \SI{10}{\micro m}$ square of the silicon nitride membrane in this step.

Based on the assumption that the experimentally observed order parameter evolution essentially follows the temperature distribution at late times, we tune the mean excitation fluence in the simulations such that the regions with temperatures $T\ge \SI{353}{K}$ reproduce the experimentally observed pattern at \SI{3141}{ns} (see Fig.~\ref{fig2}A). This leads to a fluence of \SI{1.3}{mJ/cm^2} in the simulations, while the estimated fluence in the experiments is \SI{2.6}{mJ/cm^2}. We attribute this difference to the uncertainty in determining the experimental pump spot diameter. In order to reproduce the experimental data at early times as well, we assume an instantaneous thermal equilibration of the electron temperature over the specimen thickness due to ultrafast heat transport by highly excited electrons \cite{Rettenberger1997}.

\paragraph*{Finite element simulations: Time-dependent Ginzburg-Landau approach}

Based on the ultrafast heat transfer simulations, we implement a simulation of the order parameter dynamics in the \tas{} layer using a time-dependent Ginzburg-Landau approach. We use a partial differential equation based on model ‘A’ by Hohenberg and Halperin \cite{Hohenberg1977}, describing the spatiotemporal evolution of a non-conserved order parameter:
\begin{equation*}
\dv{\phi}{t} = d\cdot\Delta\phi-\xi\cdot\pdv{F}{\phi}.
\end{equation*}

Here, $d$ is a measure of the energy related to the formation of a domain wall between NC and IC phases, and $\xi$ is a global time scale of the free-energy landscape $F$. The behavior of the free-energy functional
\begin{equation*}
F(\phi,T_\text{el}) = F_0 + \alpha_0\cdot (T_\text{el}-T_\text{C})\cdot\phi^2+\frac{\beta}{2}\cdot\phi^4+\frac{\gamma}{3}\cdot\phi^6
\end{equation*}
models a first-order phase transition in the order parameter $\phi$ as a function of the local electron temperature $T_\text{el}$  with $\alpha_0>0$, $\beta<0$, and $\gamma>0$ \cite{Toledano1987}. The critical temperature $T_\text{C}$  denotes the low-temperature side of the hysteresis loop, while
\begin{equation*}
T_1=\frac{\beta^2}{4\alpha_0\gamma}+T_\text{C}
\end{equation*}
is the high-temperature end of the bistability regime. The potential minima attributed to NC and IC phase, respectively, are as follows:
\begin{align*}
\phi_\text{NC}(T_\text{el}) &= \pm\sqrt{\frac{-\beta+\sqrt{\beta^2-4\alpha_0(T_\text{el}-T_\text{C})\gamma}}{2\gamma}} & \text{for}\quad T&<T_1,\\
\phi_\text{IC} &= 0 & \text{for}\quad T&>T_\text{C}.
\end{align*}
The bistability of $\phi_\text{NC}$ reflects the two possible alignments of the NC CDW with the lattice and the symmetry-breaking character of the phase transition. At the phase transition temperature of $T^*=\SI{353}{K}$, the free energies $F(\phi_\text{NC})$ and $F(\phi_\text{IC})$ of both CDW phases are equal:
\begin{equation*}
T^*=\frac{3\beta^2}{16\alpha_0\gamma}+T_\text{C}.
\end{equation*}
The entropy $S=-\dv*{F}{T_\text{el}}$  of both states is given by:
\begin{align*}
S_\text{NC}(T_\text{el}) &= S_0-\alpha_0\cdot\phi_\text{NC}(T_\text{el})^2 & \text{for}\quad T&<T_1,\\
S_\text{IC}(T_\text{el}) &= S_0 & \text{for}\quad T&>T_\text{C}.
\end{align*}
For the heat capacity $C = T_\text{el}\cdot(\dv*{S}{T_\text{el}})$, we find:
\begin{align*}
C_\text{NC}(T_\text{el}) &= C_0 + \frac{\alpha_0^2\cdot T_\text{el}}{\sqrt{\beta^2-4\alpha_0(T_\text{el}-T_\text{C})\gamma}} & \text{for}\quad T&<T_1,\\
C_\text{IC}(T_\text{el}) &= C_0 & \text{for}\quad T&>T_\text{C}.
\end{align*}
Due to its first-order character, the phase transition is associated with a finite latent heat $\Delta H$ and a jump in heat capacity $\Delta C$ (both evaluated at $T^*$):
\begin{align*}
\Delta C(T^*) &= C_\text{IC}(T^*) - C_\text{NC}(T^*) = \frac{2\alpha_0^2}{\beta}T^*,\\
\Delta H(T^*) &= \left[S_\text{IC}(T^*) - S_\text{NC}(T^*)\right]\cdot T^* = -\frac{3}{4}\frac{\alpha_0\beta}{\gamma}T^*.
\end{align*}
By normalizing the NC CDW potential minima to $\phi_\text{NC}(T_\text{RT}) = 1$ at room temperature, and using literature values for $\Delta C$ \cite{Suzuki1985} and $\Delta H$ \cite{Wilson1975}, we can derive values for the constants $\alpha_0$, $\beta$, and $\gamma$:
\begin{align*}
\alpha_0 &= \SI{8.04}{J/mol/K},\\
\beta &= \SI{-151}{J/mol},\\
\gamma &= \SI{626}{J/mol}.
\end{align*}
This leads to a width of hysteresis $\Delta T = T_1-T_\text{C} = \SI{1.1}{K}$ which is very well compatible with our upper estimate of \SI{4}{K} (see main text).

In our simulations, the parameter $d$ effectively prevents the formation of step-like phase boundaries between NC and IC CDW regions. Given the spatial resolution in measuring the phase boundary width in the steady-state experiment, we deduce an upper limit of $d = \SI{1}{nm^2/ps}$. Smaller values of $d$ only have a moderate effect on the ultrafast order parameter dynamics on the length scales considered in this work, such that the specific selection of $d$ is not of great significance. We choose a value of $d = \SI{0.1}{nm^2/ps}$ in our simulations (cf. Fig.~\ref{figS8}A). The global time scale $\xi$ is the only remaining parameter in order to fit the magnitude of the initial order parameter suppression during optical excitation. A value of $\xi = \SI{5e-4}{mol/J/ps}$ is used in the simulations presented in Fig.~\ref{fig4}.

Initially, the simulation volume is in the room-temperature potential minimum of the NC phase with $\phi_\text{NC}(T_\text{RT}) = 1$. In the experiments, the confinement of the optical excitation to the field of view of the electron beam and the good coupling to the heat bath suppress independent nucleation of the NC phase. The chiral state of the re-established NC phase is therefore determined by the initial NC orientation. Generally, the presence of the second NC potential minimum would allow for modeling of the probabilistic appearance of “mirror domains” \cite{Zong2018}. However, the deterministic nature of our simulation prohibits this kind of relaxation dynamics in agreement with our experimental observations.

\subsection*{Supplementary text}

\paragraph*{Structured excitation profile inside the circular gold aperture}
\label{supptext:excitation}

In the main text, we state that the precise shape of the (almost) circular aperture on top of the specimen determines the spatial profile of the excitation density and gives rise to a four-lobe excitation pattern as shown in Fig.~\ref{fig2}E. This behavior arises due to the symmetry breaking by the linear optical polarization. In order to further support this description, we present an additional dataset at a fixed pump/probe delay and pump fluence, but with a varying angle of linear polarization between \SI{0}{\degree} and \SI{45}{\degree} (see Fig.~\ref{figS9}). Comparing the images at \SI{0}{\degree} and \SI{45}{\degree} linear polarization, one can see that a part of the four-lobe pattern is rotated by \SI{45}{\degree}.

Some of the additional complexity that is visible in the pattern at \SI{45}{\degree} arises from the fact that the polarization rotation is done outside of the column of the electron microscope. After polarization manipulation, the beam passes additional mirrors with a combination of s- and p-polarization. This leads to an additional phase-shift per reflection between s- and p-components of the beam, and thus a deviation from linear towards elliptic polarization characteristics.

\paragraph*{Orientation of NC/IC phase boundaries along preferential directions}
\label{supptext:orientation}

In the steady-state experiment, we observe the formation of NC/IC domain walls along preferential directions that reflect the hexagonal symmetry of the underlying lattice. In order to analyze these preferential directions, we first determine the relative rotation between diffraction and imaging modes of our TEM with the help of an anisotropically etched silicon frame. Most of the rotation is already compensated in the imaging system of the TEM, so that the residual offset only amounts to \SI{1.2(4)}{\degree}.

From the steady-state DF image series, we select four images that show the most pronounced formation of straight phase boundaries. After exclusion of those phase boundaries that seem to exhibit small “steps”, we extract the angles (modulo \SI{60}{\degree}) of the phase boundaries indicated in Fig.~\ref{figS10}A. The resulting angle interval is drawn in gray over the diffraction image of the NC phase in Fig.~\ref{figS10}B. The obtained direction seems to indicate some correspondence with the \hkl{1 1 -2 0} planes of the underlying hexagonal lattice, but future work will be required to associate this with a particular microscopic origin.

\clearpage
\subsection*{Supplementary figures}

\begin{figure}[!h]
\centering
\includegraphics{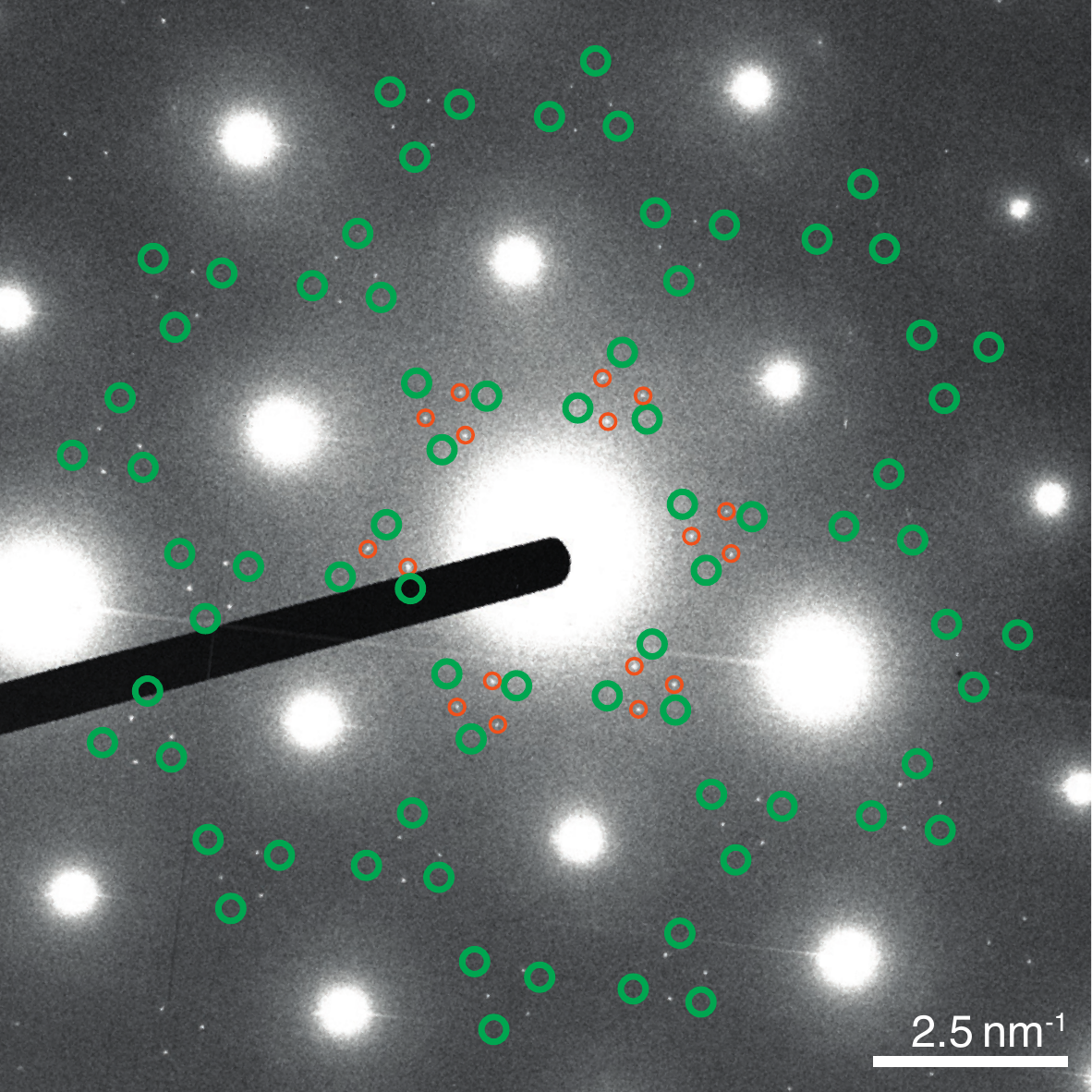}
\caption{
\textbf{Diffraction pattern of the IC phase of \tasbold{}.}\\
Second-order IC superstructure reflections are visible along with bright structural reflections and some inelastic background. As in the NC phase, first-order reflections are forbidden due to stacking periodicity \cite{Scruby1975}. For clarity, those IC reflections that appear closest to the direct beam have been highlighted by orange circles. Position and size of the individual apertures in the DF aperture array are indicated by green circles (cf. Fig.~\ref{fig1}B). The superstructure peaks of the IC phase appear close to the positions of the individual apertures in the DF array, however, there is no overlap between them.
}
\label{figS1}
\end{figure}
\clearpage

\begin{figure}[!p]
\centering
\includegraphics[width=\textwidth]{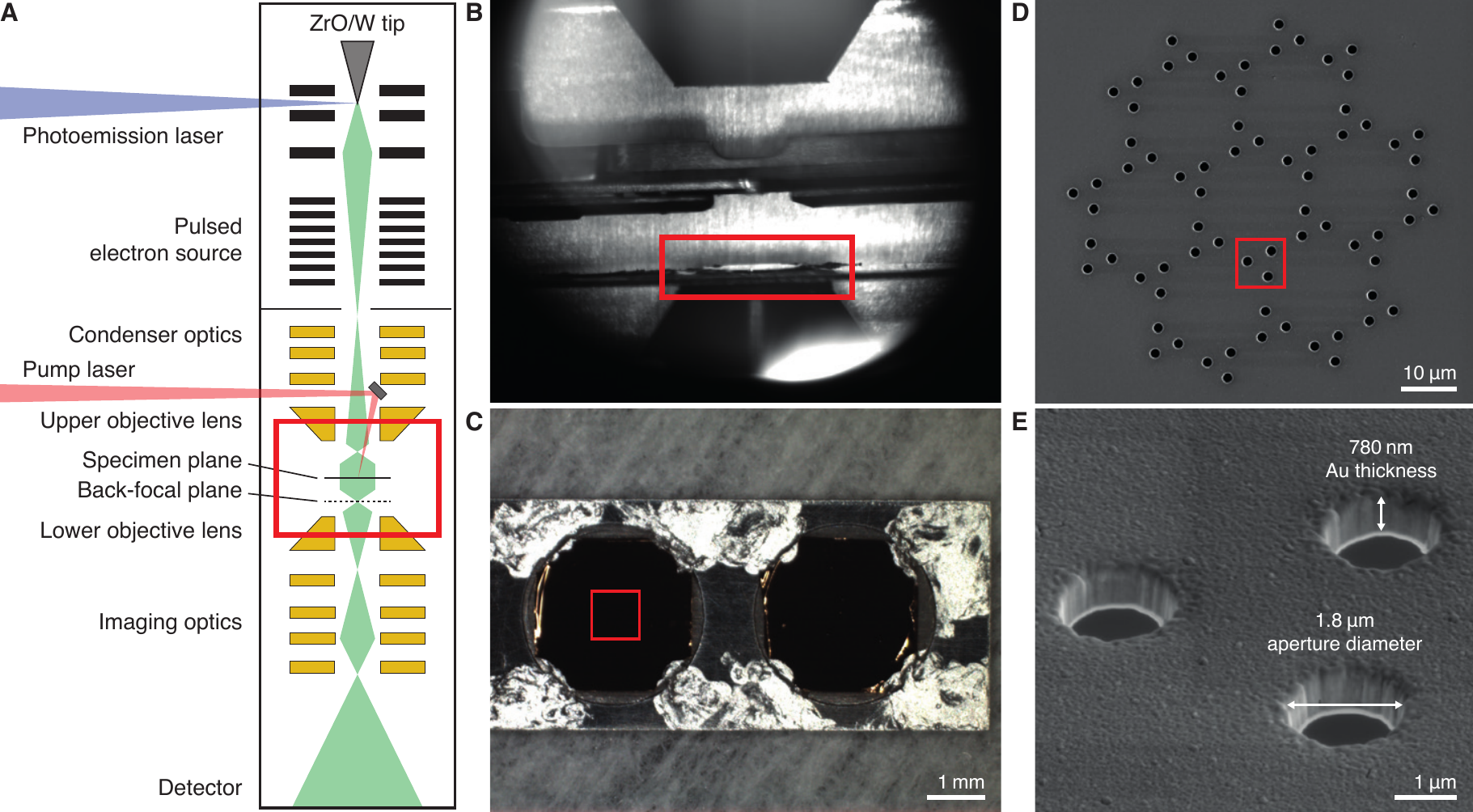}
\caption{
\textbf{Extended schematic of the experiment and the DF aperture array.}\\
(\textbf A)~Schematic of the Göttingen Ultrafast Transmission Electron Microscope (UTEM) in the configuration used in this study. Red boxes indicate the field of view of the subsequent subfigure.
(\textbf B)~Optical camera image of the objective lens gap. Specimen holder and custom aperture holder are visible in the gap. The \SI{3}{mm} DF aperture frame is marked.
(\textbf C)~Two DF aperture frames installed in the aperture holder.
(\textbf D)~Scanning electron micrograph of the DF aperture array taken after ion milling of the hole pattern.
(\textbf E)~Cross-sectional view of a hole triplet (\SI{52}{\degree} specimen tilt; image taken on another DF aperture array from the same batch).
}
\label{figS2}
\end{figure}
\clearpage

\begin{figure}[!p]
\centering
\includegraphics[width=\textwidth]{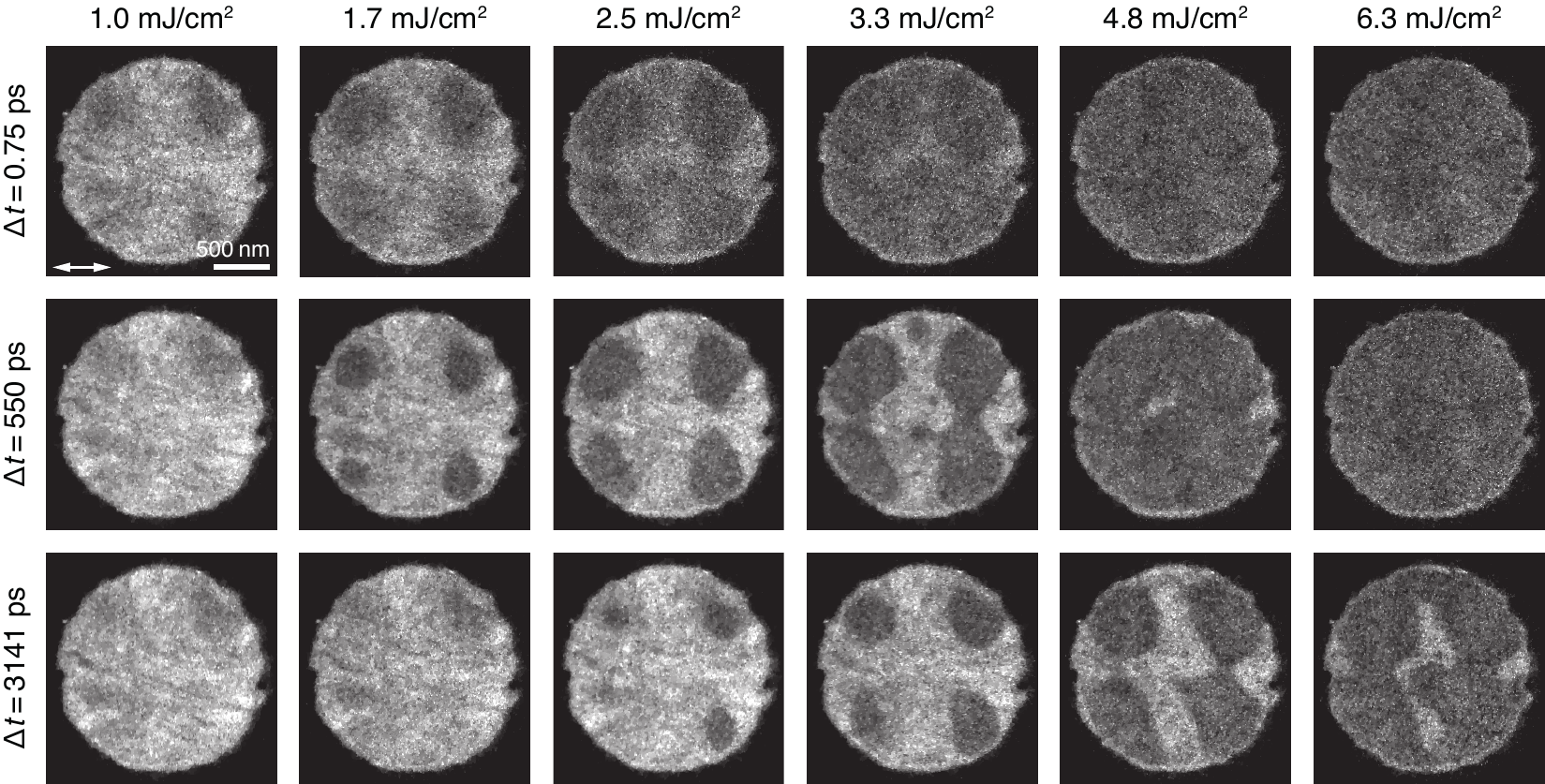}
\caption{
\textbf{Ultrafast DF images as a function of laser fluence at fixed pump/probe delays.}\\
Images taken with horizontal pump polarization as indicated by white arrow (\SI{11}{min} integration time per frame at \SI{0.75}{ps} delay, \SI{5.5}{min} integration time at \SI{550}{ps} and \SI{3141}{ps}). The spatial profile of the excitation density has been extracted from this dataset (see subsection “\nameref{mat:excitation}”).
}
\label{figS3}
\end{figure}
\clearpage

\begin{figure}[!p]
\centering
\includegraphics{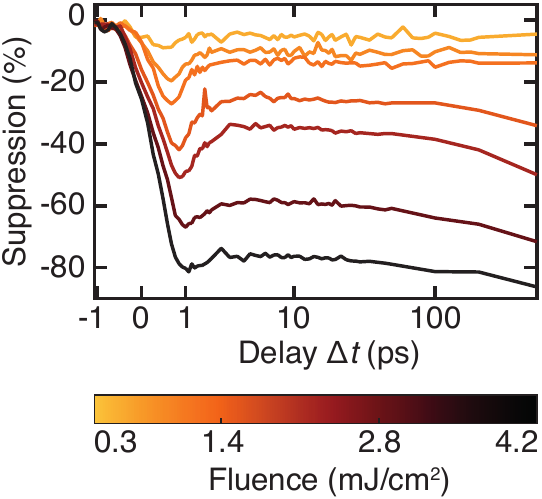}
\caption{
\textbf{Fluence-dependent ultrafast transmission electron diffraction.}\\
Delay curves of spatially averaged NC CDW diffraction spot intensities, illustrating the transition from low-fluence transient NC CDW suppression to high-fluence IC phase formation (cf. green and orange curves in Fig.~\ref{fig2}C, bottom).
}
\label{figS4}
\end{figure}
\clearpage

\begin{figure}[!p]
\centering
\includegraphics[width=\textwidth]{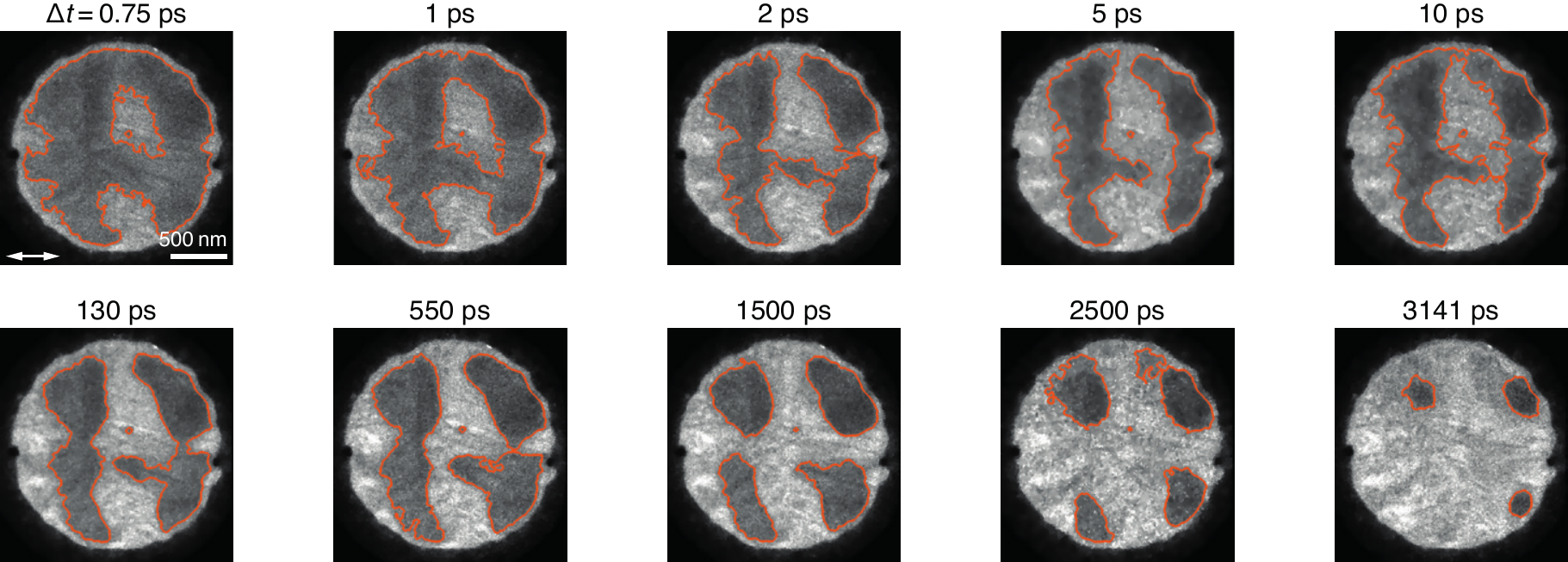}
\caption{
\textbf{Demonstration of image segmentation.}\\
Demonstration of the image segmentation approach based on the image series presented in Fig.~\ref{fig2}A. The contrast of the additional images at 5, 10 and \SI{2500}{ps} has been adjusted in order to accommodate for a different intensity ratio of bright and dark regions (due to a slight drift of the DF aperture array during image acquisition).
}
\label{figS5}
\end{figure}
\clearpage

\begin{figure}[!p]
\centering
\includegraphics[width=\textwidth]{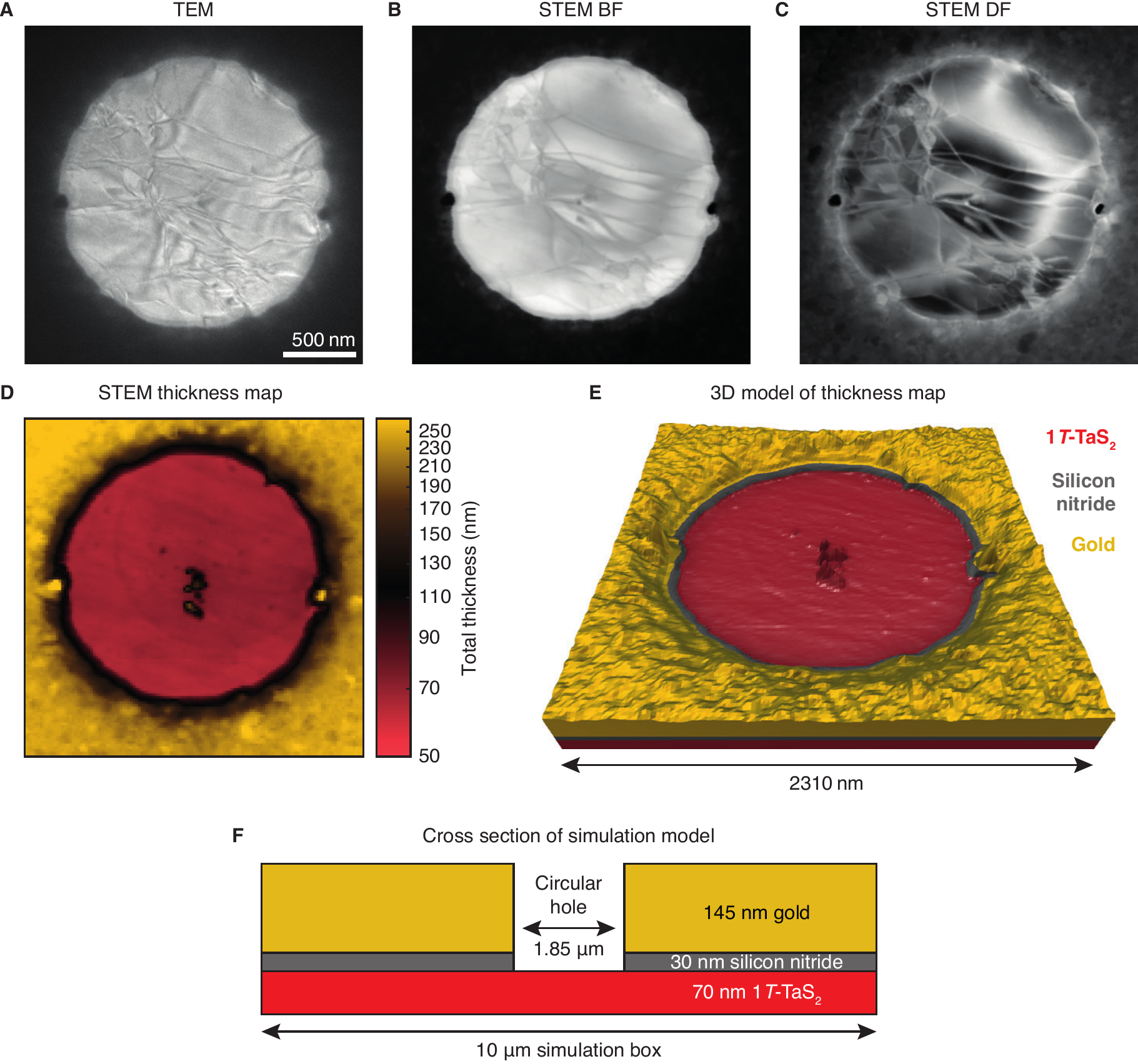}
\caption{
\textbf{Imaging of the specimen structure with a continuous electron beam (TEM imaging at $\mathbf{120}\;\text{kV}$ acceleration voltage; STEM imaging at $\mathbf{200}\;\text{kV}$ acceleration voltage).}\\
(\textbf A)~TEM image.
(\textbf B)~STEM bright-field (BF) image.
(\textbf C)~STEM annular DF image (i.e., using an annular electron detector).
(\textbf D),~(\textbf E)~Specimen thickness map derived from STEM-EELS data. Different specimen components are indicated by different colors.
(\textbf F)~Cross-section of the simulation model used in the finite element simulations.
}
\label{figS6}
\end{figure}
\clearpage

\begin{figure}[!p]
\centering
\includegraphics{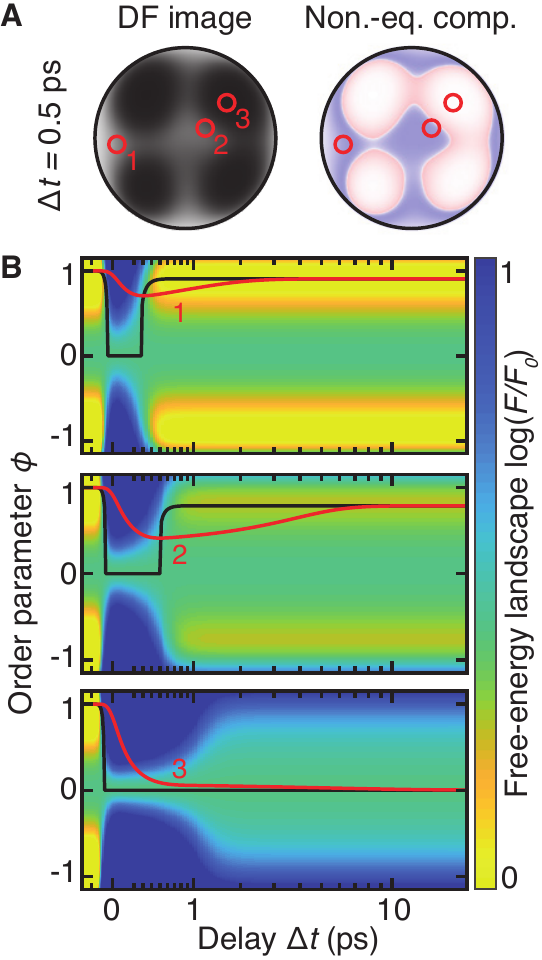}
\caption{
\textbf{Order parameter trajectories on the free-energy surface for different levels of local excitation.}\\
(\textbf A)~Representative locations with light NC CDW suppression~(1), strong suppression~(2), and IC phase formation~(3) (cf. Fig.~\ref{fig4}, A and C).
(\textbf B)~Order parameter trajectories at the locations indicated in A. Red curve: Transient value of order parameter. Black line: Global free-energy minimum as a function of delay $\Delta t$ (shown for positive values of the order parameter only).
}
\label{figS7}
\end{figure}
\clearpage

\begin{figure}[!p]
\centering
\includegraphics{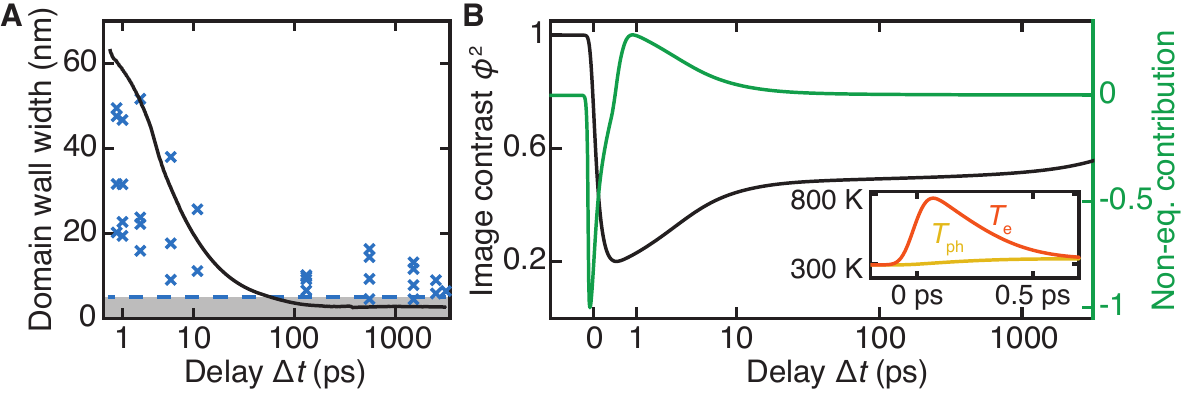}
\caption{
\textbf{Domain wall width and extended simulation results.}\\
(\textbf A)~Width of NC/IC domain walls as extracted from time-resolved experimental data at different positions (blue data points; cf. Fig.~\ref{fig2}A), and simulation results (black line; cf. Fig.~\ref{fig4}A). The dashed blue line is the domain wall width extracted from the steady-state images (cf. Fig.~\ref{fig3}A). The shaded region of the plot indicates the resolution limit of the experiment.
(\textbf B)~Spatially averaged delay curves corresponding to the simulation results in Fig.~\ref{fig4}, A and C (black and green curves, respectively). Inset: Electron (orange) and lattice temperatures (yellow) during electron-lattice equilibration.
}
\label{figS8}
\end{figure}
\clearpage

\begin{figure}[!p]
\centering
\includegraphics{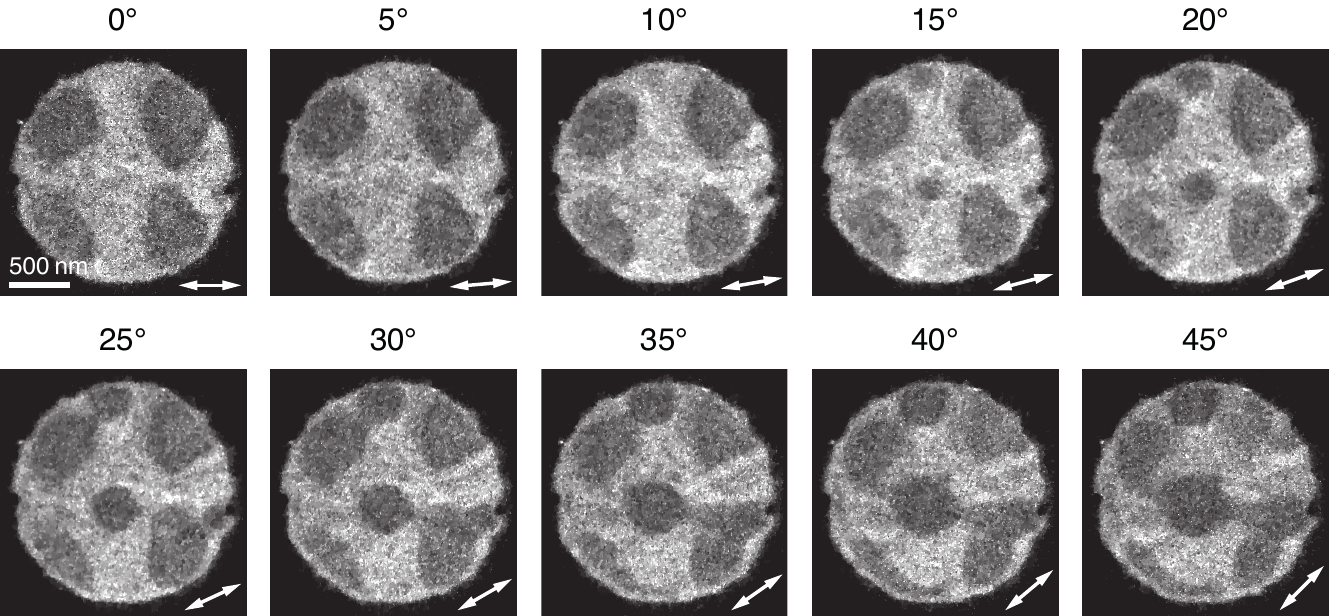}
\caption{
\textbf{Ultrafast DF images as a function of linear pump polarization.}\\
Pump polarization indicated by white arrows. Images taken at a fixed \SI{550}{ps} pump/probe delay and \SI{2.6}{mJ/cm^2} pump fluence (\SI{11}{min} integration time per frame). This figure is discussed in the supplementary text (see subsection “\nameref{supptext:excitation}”).
}
\label{figS9}
\end{figure}
\clearpage

\begin{figure}[!p]
\centering
\includegraphics{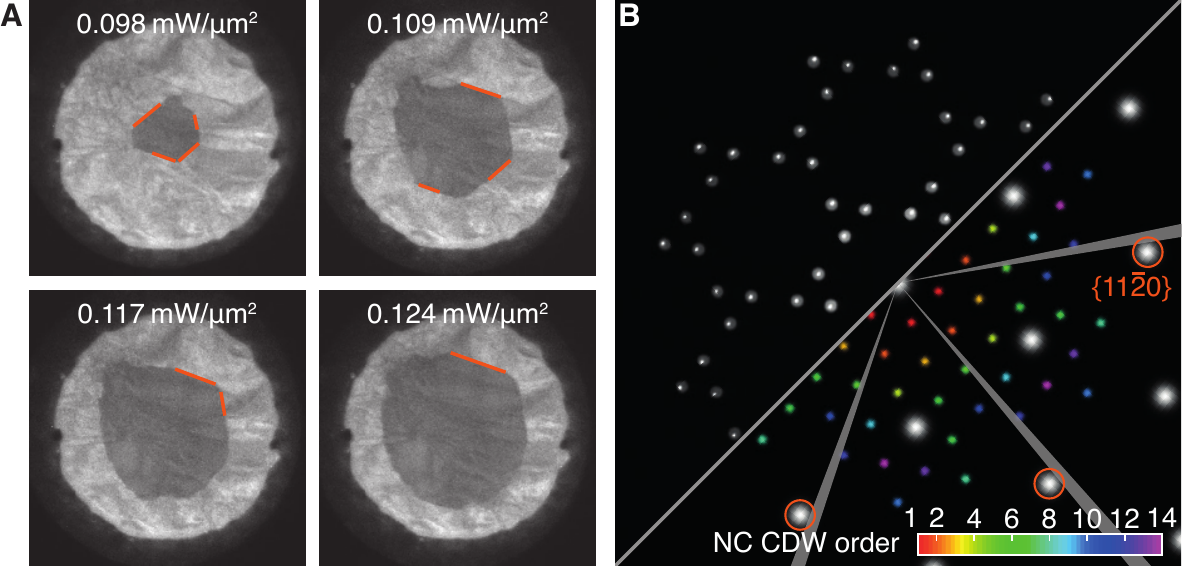}
\caption{
\textbf{Identification of reciprocal lattice directions in the steady-state DF images.}\\
(\textbf A)~Selection of steady-state DF images that exhibit pronounced formation of NC/IC phase boundaries along preferential lattice directions. Orange lines indicate the subset of boundaries that has been used to determine the angular intervals shown in B.
(\textbf B)~Top half: Diffraction image of the NC phase with the DF aperture array (only second-order satellites are visible). Bottom half: Diffraction pattern generated from theoretical spot positions of the underlying lattice (white spots) and of NC CDW orders from 1 to 14 that belong to the \hkl(0000) structural reflection (rainbow color scale) \cite{Spijkerman1997}. Grey angle intervals indicate the reciprocal lattice directions corresponding to the lattice planes indicated in A. Both diffraction images have been aligned to precisely reflect the orientation of the real-space images in A. This figure is discussed in the supplementary text (see subsection “\nameref{supptext:orientation}”).
}
\label{figS10}
\end{figure}
\clearpage

\subsection*{Supplementary tables}

\begin{table}[!h]
\centering
\small
\begin{tabular}{ |c|c|c|c|c| } 
\hline
\textbf{Material} & \textbf{\makecell{Complex \\ refractive index}} & \textbf{\makecell{Density and \\ molar mass}} & \textbf{\makecell{Thermal \\ conductivity}} & \textbf{\makecell{Specific heat \\ capacity}}\\
\hline
\textbf{Gold} & Ref.~\cite{Johnson1972} & \cite{Haynes2017} & \cite{Haynes2017} & \cite{Haynes2017} \\
\hline
\textbf{\makecell{Silicon nitride \\ membrane}} & \cite{Luke2015} & \cite{Norcada2017} & \cite{Jain2008} & \cite{Jain2008} \\
\hline
\textbf{1\textit{T}-TaS$_\mathbf2$} & \cite{Beal1975} & \cite{Haynes2017} & \cite{Nunez-Regueiro1985} & \cite{Meyer1975,Suzuki1985} \\
\hline
\end{tabular}
\caption{
\textbf{Physical material properties used in the finite element simulations.}
}
\label{tabS1}
\end{table}

\subsection*{Supplementary movies}

\paragraph*{\small\textbf{Movie S1. Ultrafast DF image series.}}\mbox{}\\
\small Data taken with \SI{2.6}{mJ/cm^2} pump fluence, \SI{11}{min} integration time per frame, horizontal pump polarization. This dataset is used to extract the delay curves shown in Fig.~\ref{fig2}C.
\label{movieS1}

\paragraph*{\small\textbf{Movie S2. Ultrafast bright-field (BF) image series.}}\mbox{}\\
\small Data taken under the same experimental conditions as Movie~\hyperref[movieS1]{S1} (except for a shorter \SI{5.5}{min} integration time per frame). Only the DF aperture array is retracted from the column of the microscope. Note that no CDW contrast is visible, and that the movement of diffraction contrast lines indicates a mechanical oscillation of the membrane on nanosecond timescales, however, with no discernible influence on the DF images (cf. Movie~\hyperref[movieS1]{S1}).
\label{movieS2}

\paragraph*{\small\textbf{Movie S3. Steady-state DF image series.}}\mbox{}\\
\small Data taken with \SI{5}{s} integration time. A subset of these images is also presented in Fig.~\ref{fig3}A.
\label{movieS3}

\paragraph*{\small\textbf{Movie S4. Time-dependent Ginzburg-Landau simulation of ultrafast DF images.}}\mbox{}\\
\small A subset of these images is also presented in Fig.~\ref{fig4}A.
\label{movieS4}

\paragraph*{\small\textbf{Movie S5. Time-dependent Ginzburg-Landau simulation (non-equilibrium contribution).}}\mbox{}\\
\small A subset of these images is also presented in Fig.~\ref{fig4}C.
\label{movieS5}

\end{document}